\documentclass[10pt,a4paper]{article}
\usepackage{dsfont,amsmath,amssymb,graphicx,cite}

\begin{document}

\noindent{\Large\textbf{Cryptographic key distribution over a public network via variance-based watermarking in compressive measurements}}

\vspace{5pt}
\noindent\textbf{Wen-Kai Yu$^*$}

{\raggedright\footnotesize\textit{Center for Quantum Technology Research, School of Physics, Beijing Institute of Technology, Beijing 100081, China}}

\renewcommand{\thefootnote}{\fnsymbol{footnote}}
\footnotetext[1]{\footnotesize{Corresponding author.

\ \ E-mail address: yuwenkai@bit.edu.cn (W.-K. Yu).}}

\vspace{8pt}
\noindent{\small A B S T R A C T}

\noindent\rule{\textwidth}{0.05em}

\noindent{\small The optical communication has an increasing need for security in public transmission scenarios. Here we present a protocol for cryptographic key distribution over a public network via photon-counting compressive imaging system with watermarking, which utilizes watermarking technique to distribute secure keys, and uses reconstructed images for simultaneous identity authentication and tampering identification. The watermark is embedded in the rearranged compressed measurements of the object, and then the signal is transmitted through a public network. At the receiving terminal, the legitimate users can easily extract the watermark as the cryptographic key by using initial keys and the variance characteristic of random measurements. Artificial tampering and attacks can be detected by the accurately retrieved images. The realization of this protocol is a step forward toward the practical applications, and will be beneficial for the broader fields of optical security in many ways.}

\noindent\rule{\textwidth}{0.05em}



\noindent\rule{\textwidth}{0.05em}

\section{Introduction}
In recent years, digital information has been widely transmitted through insecure public network \cite{Lian2009} where the data is particularly easy to be wiretapped by others, thus the techniques for information security are of great importance. Many optical encryption methods have been developed. As far as we know, the double random phase encoding (DRPE) technique \cite{Javidi1995} proposed in 1995 may be the first successful example in optical encryption. Based on optical parallel characteristics, it generally uses double spatial light modulators for phase encoding and array charge-coupled devices (CCDs) for recording the intensities \cite{Sheridan2003,Sheppard2010,Lancis2010,Situ2017}. Lately, the DRPE has been integrated with photon-counting to obtain better security and verification performance \cite{Javidi2011,Rajput2014,Maluenda2015}, but its sampling is still redundant. Fortunately, the newly developed theory of compressed sensing (CS) \cite{Donoho2006,Candes2006,Candes2008} provides a new technical solution for DRPE compression during the optical sampling \cite{RLi2015}, and allows one to accurately reconstruct the signal known \textit{a priori} to be sparse or compressive from its linear projections, with a sampling rate far below the Nyquist rate \cite{Donoho2006,Candes2006}. Given all these advantages of CS, many methods in optical security field, like Ptychography \cite{Lee2015}, quick response codes \cite{Wang2015}, have been recently combined with compressed DRPE to enhance their performance. On the other hand, CS has developed a single-pixel camera \cite{Baraniuk2008,CBLi2010} to acquire wider applications, like in remote sensing \cite{Zhao2012,YuSR2014}, adaptive ghost imaging \cite{Yu2014}, biological imaging \cite{Yu2016}, phase retrieval \cite{Yu2017}, and so forth. Recent works have successfully combined single-pixel compressive imaging with optical encryption, where the signal and its compressed measurements are treated as the plaintext and the ciphertext. This technology is proven to be able to increase the security \cite{Orsdemir2008,DXiao2015} and thus becomes popular.

With the developments of information age, one important issue is how to improve the security of the information delivered by the public network, especially in financial transactions, secure video communication, Internet conferences, telemedicine, and the like. Our earlier works have proposed a protocol for high-speed secure key distribution over a public network \cite{ShenLi2013,Yu2013}, in which the legitimate parties acquire the cryptographic keys from the images recovered by computational ghost imaging algorithms. Note that the watermarking \cite{Huang2014,Liu2016,Weng2016} and data hiding \cite{XZhang2011,MLi2016} are two representative choices for privacy protection \cite{Xia2016} by embedding a label to a visible image, we find that they may be useful for improving the performance of the secure key distribution. However, to our knowledge, most of traditional watermarking techniques needed to embed the watermark into the non-overlapping blocks of original image \cite{Weng2016,XZhang2011,MLi2016}, and failed to consider authentication and tampering identification aspects. Besides, the watermarked image transmission may incur the content leakage.

In this work, we propose a watermarking-based protocol for cryptographic key distribution over a public network, in which the watermark is regarded as the cryptographic key to be distributed. In the carrier generation process, based on a single-pixel compressive imaging system, the server uses a digital micromirror device (DMD) for random modulation and a photomultiplier tube (PMT) for photon counting. Then the server will embed the watermark in the rearranged compressed measurements which have no spatial correlation, rather than in the image or its non-overlapping divided blocks, thus increasing the confidentiality and imperceptibility of the cryptographic key, and without the content leakage. After that, the disordered sequence will be sent to the legitimate users through the public network. With the aid of the initial keys and the variance characteristic of measurements, the valid parties can exactly extracted the distributed cryptographic key. By utilizing CS algorithm, the users can reconstruct the images for simultaneous efficient authentication and tampering identification. Additionally, unlike traditional optical communication schemes that are mainly based on the two-dimensional encrypted images, our protocol also offers an alternative approach for embedding the watermark directly in the randomly-ordered one-dimensional sampled signal as well as taking full advantage of the variance characteristic of measurements, which will have a great impact on research in the broader fields of optical security, especially in cryptographic key distribution applications.

\section{Protocol}
Our protocol applies CS for carrier generation and authentication. In CS, an image signal $x$ of pixel size $s\times t$ can be reshaped into a one-dimensional column vector of length $N=s\times t$. Every pattern encoded on the DMD has the same size as the object image, and is flattened accordingly into a row vector $a_j$ of length $N$. There are $M$ random binary patterns that will be sequentially fed into the DMD, the row vectors reshaped from these $M$ patterns are rearranged into a measurement matrix $A\in\mathds{R}^{M\times N}$. Rather than directly sampling the image pixel values, CS captures the linear projections of the signal at a rate that is significantly below the Nyquist rate, namely, it samples the inner product between $M$ modulating row vector $a_j$ and the signal $x$. Thus the problem can be given by a set of linear equations: $y=Ax+e$, where $y$ is the observed measurement vector, $e\in\mathds{R}^{M\times 1}$ denotes the stochastic noise. Hence simultaneous compression and sampling of an image can be achieved. Generally, the natural signal $x$ can be sparsely represented in a certain basis (e.g., discrete cosine transform basis, Fourier transform basis, wavelet basis) $\Psi=[\psi_1,\psi_2,...,\psi_N]$, thus $x=\Psi x'$, or $x=\sum\limits_{i=1}^N{{x'}_i}{\psi_i}$, where $x'\in \mathds{R}^{N\times 1}$ is the coefficient sequence of $x$. Since the total variation (TV) regularization can ensure excellent reconstructed image quality by preserving the edges accurately \cite{CBLi2010}, here we use TV solver as the sparse basis. In order to satisfy the restricted isometry property (RIP) \cite{Candes2008}, the column vectors taken from arbitrary subsets of the measurement matrix $A$ should be approximately orthogonal and incoherent with the sparse basis. It is found that a completely random matrix $A$ works well in general and is largely incoherent with any fixed basis. Therefore, the random measurement matrix is adopted here, and its randomness can render the measurement data unintelligible. The object image can be reconstructed precisely from the dimension reduction measurements via optimization algorithm.
\begin{figure}[htbp]
\centering
\includegraphics[width=0.95\linewidth]{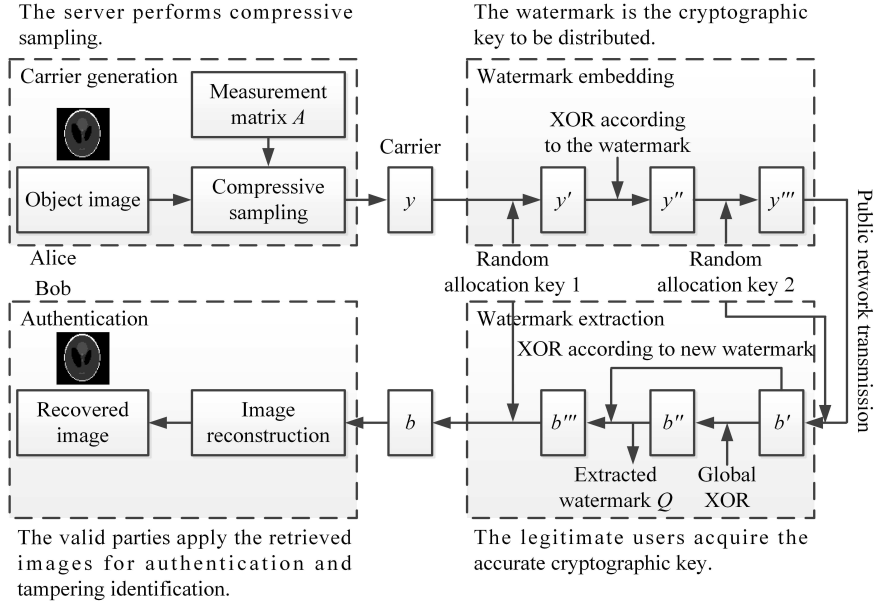}
\caption{Schematic diagram of the protocol based on watermarking for cryptographic key distribution over a public network.}
\label{fig:protocol}
\end{figure}

\begin{figure}[htbp]
\centering
\includegraphics[width=0.95\linewidth]{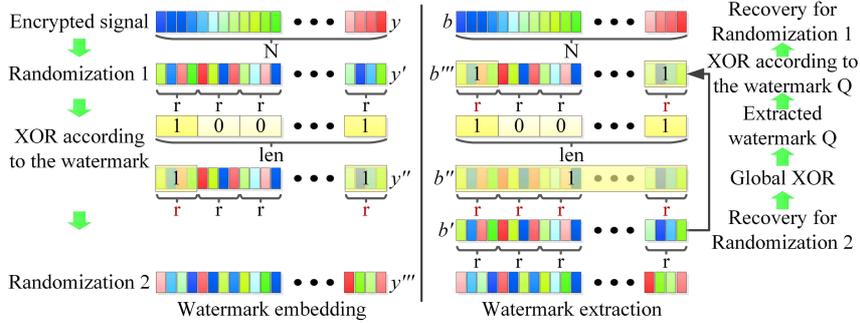}
\caption{Procedure for the watermark embedding (left panel) and extraction (right panel).}
\label{fig:watermark}
\end{figure}

This protocol is an extension of our previous work \cite{ShenLi2013,Yu2013}. Here, the measurement matrix is used as a key known by the legitimate parties in advance, and a watermark label is embedded to the measured signal for cryptographic key distribution. As depicted in figure~\ref{fig:protocol}, our protocol can be divided into four parts: carrier generation, watermark embedding, watermark extraction, and authentication. Figure~\ref{fig:watermark} shows the detailed procedure for watermark embedding and extraction.

\subsection{Carrier generation}
In the carrier generation step, the server performs compressive sampling \cite{Candes2006} via a single-pixel imaging system, which is simpler than DRPE setup. The object and the measurement matrix can be denoted as $x$ and $A$, then the carrier $y\in\mathds{R}^{M\times 1}$ can be acquired by the linear equations $y=Ax+e$. For different legitimate parties, the object $x$ and the measurement matrix $A$ can be different. Therefore, our protocol can realize multiparty communication.

\subsection{Watermark embedding}
Here the watermark is treated as the cryptographic key to be distributed. In the watermark embedding process as shown in the left half of figure~\ref{fig:watermark}, $y$ will be disordered by the server via the random allocation key 1 to obtain $y'$, which is then divided into $M/r$ (the same as the length of the watermark) groups with each of size $r$. For the $i$th group, if the element value $w_i$ in the watermark equals to 1, then all the bits in the corresponding group (denoted as $seq_i$, $i=1,2,\cdots,M/r$) of $y'$ are exclusive-ORed (XOR), otherwise, leaving it as it is, i.e.,
\begin{equation}
seq_i=\left\{\begin{array}{l}
\textrm{XOR}(seq_i),\ \textrm{if}\ w_i=1;\\
seq_i,\ \textrm{otherwise}.
\end{array}\right.
\label{eq:embedding}
\end{equation}
Here using XOR for watermark embedding is a common operation \cite{XZhang2011}. By this means, we will get the sequence $y''$, which will be disordered again according to the random allocation key 2 to obtain $y'''$. After that, the signal $y'''$ will be transmitted through the public network to the legitimate users.

\subsection{Watermark extraction}
In the watermark extraction as shown in the right half of figure~\ref{fig:watermark}, the legitimate receiver restores $b'$ via the random allocation key 2, and then carries out XOR operation on every bit of $b'$ to obtain $b''$ (this operation we call global XOR). The values $w'_i$ of the watermark bits are extracted via
\begin{equation}
w'_i=\left\{\begin{array}{l}
1,\ \textrm{if}\ \textrm{var}(b''_i)<\textrm{var}(b'_i);\\
0,\ \textrm{otherwise},
\end{array}\right.
\label{eq:extraction}
\end{equation}
where $\textrm{var}$ denotes the variance. According to the extracted watermark and the group assignment, the XOR operation of the whole bits in the divided groups of $b'$ that correspond to the extracted watermark bit value 1 is performed, also via Eq.~\ref{eq:embedding}, to get $b'''$. In theory, $b'$ and $b'''$ equals to $y''$ and $y'$, respectively. The correct sequence $b$ is then recovered following the random allocation key 1. Theoretically, $b$ equals to $y$ with a great probability. Through the above operations, legitimate users will finally get the corresponding cryptographic keys distributed from the server. The cryptographic keys can be used to encrypt the plaintext and decrypt the ciphertext during the multiparty communication.

\subsection{Authentication}
Finally, the receiving parties use the TVAL3 algorithm \cite{CBLi2010} to recover the image from $A$ and $b$ for identity authentication, as demonstrated in figure~\ref{fig:protocol}. The core TV norm minimization of TVAL3 can be described as:
\begin{equation}
\mathop{\min}\limits_{x}\mathop{\sum}\limits_{i}{\left\|{D_ix}\right\|_p}+\frac{\mu}{2}\left\|{b-Ax}\right\|_2^2,\ \textrm{s.t.}\ Ax+e=b,\ D_ix=x'_i\ \textrm{and}\ x\geq0\ \textrm{for\ all}\ i,
\label{eq:TVnorm}
\end{equation}
where ${\left({\left\|u\right\|}_p\right)^p}={\sum\nolimits_{i=1}^N{\left|{u_i}\right|}^p}$ stands for $l_p$ norm. ${D_i}x$ is the discrete gradient vector of $x$ at position $i$, $D$ is the gradient operator, and $\mu>0$ is a penalty constant scalar used to balance these two terms. $\sum\nolimits_i{{{\left\|{{D_i}x}\right\|}_p}}$ is the TV regularization term, which is isotropic for $p=2$, and anisotropic for $p=1$. Here we use the isotropic (the higher, 2nd-order) derivative to get better image quality for authentication.

\section{Simulation}
To obtain a quantitative measure of the image quality, the peak signal-to-noise ratio (PSNR) is defined here as a figure of merit:
\begin{equation}
\textrm{PSNR}=10\log(255^2/\textrm{MSE}),
\label{eq:PSNR}
\end{equation}
where $\textrm{MSE}=\frac{1}{st}\sum\nolimits_{i,j=1}^{s,t}[U_o(i,j)-\tilde U(i,j)]^2$. The MSE describes the squared distance between the recovered image $\tilde U$ and the original image $U_o$ for all $s\times t$ pixels. Naturally, the larger the PSNR value, the better the quality of the image recovered.

Additionally, we introduce another unitless performance measure, the bit error rate (BER), which is defined as
\begin{equation}
BER=\frac{1}{{len}}\sum\limits_{i = 1}^{len} {\left| {{W_i} - {Q_i}} \right|}.
\label{eq:BER}
\end{equation}
Here, $W$ is the original watermark, $Q$ is the extracted watermark, and $len$ is the length of $W$. Therefore, the BER is the number of bit errors divided by the total length of the watermark bits during a time interval studied. The generation of error bits is mostly due to noise, randomness, or tampering.
\begin{figure}[htbp]
\centering
\includegraphics[width=0.65\linewidth]{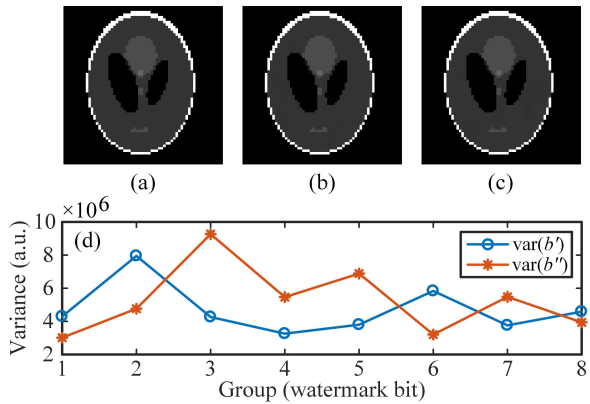}
\caption{Simulation results. (a) is the original phantom image of $64\times 64$ pixels; (b) and (c) are the recovered results via TVAL3, with a BER of 0.00\% and 0.00\%, and a PSNR of 64.72~dB and 62.48~dB, respectively. The length of each watermark for (b) and (c) is 8~bits (1~bytes) and 64~bits (8~bytes), respectively. In order to extract the watermark more accurately, the same watermark repeats five times. The number of measurements $M$ (i.e., the length of the measured signal) for (b) and (c) is 1280 and 5120, thus the signal is distributed into 40 and 320 groups (i.e., $r=32$ and $16$~bits per groups), respectively. (d) shows the variance of the first eight groups of $b'$ and $b''$ of (b). The watermark bit values are extracted to be 1 for $\textrm{var}(b'')<\textrm{var}(b')$, 0 otherwise. Since the original watermark used here is ``1 1 0 0 0 1 0 1'', the extracted watermark is correct.}
\label{fig:simulation}
\end{figure}
\begin{figure}[htbp]
\centering
\includegraphics[width=0.75\linewidth]{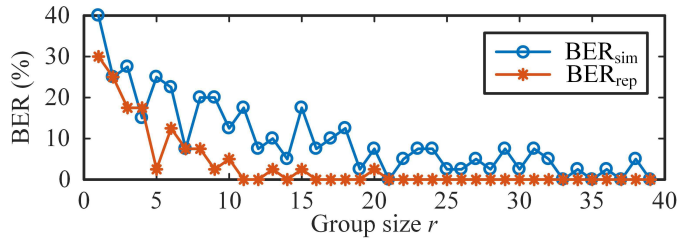}
\caption{BER of the non-repeating watermark and the one with five repeats versus the group size $r$. The original image is the same with figure~\ref{fig:simulation}(a), and the watermark has 40~bits (5~bytes).}
\label{fig:BERsimrep}
\end{figure}

To illustrate our strategy, the simulation procedure is presented in figure~\ref{fig:simulation}. After the sampling, the watermark embedding and extraction will be performed. According to the CS theory, the original image can be any complex gray-scale or colour-scale image, as long as it is sparse or compressive. Generally speaking, many natural images except pure noisy ones have sparse representations when they are expressed in a certain basis. It is worth noting that the watermark can be not only a binary vector but also a gray-scale image, because the image pixel values can be easily converted into a binary sequence and linked end-to-end to form a long one in watermark embedding, and vice versa during watermark extraction. After that, the CS image [figure~\ref{fig:simulation}(b) and (c)] can be recovered accurately if and only if the extracted watermark is exactly consistent with the original one. Although the poor quality of the recovered image is possibly caused by the measurement noise, we find that repeatedly using the same watermark can greatly improve the image quality. Here the watermark is reused five times in one set of measurements to reduce the negative effect of the measurement noise on the quality of the extracted watermark. The number five used here is an empirical parameter. Actually, for a 40-bit watermark with five repetitions, when the group (block) size $r$ is set to 11, the BER is already low enough; when $r$ is larger than 20, the watermark is extracted exactly with high immunity to noise, as illustrated in figure~\ref{fig:BERsimrep}. Therefore, by applying this strategy, the quality of the CS image can be used for authentication to validate whether there exists an attacker or not. The demonstrating experiment is performed by increasing the measurement number. There is a tradeoff between the group size for the BER performance and the length of the watermark.

\section{Experimental validation and results}
\begin{figure}[htbp]
\centering
\includegraphics[width=0.7\linewidth]{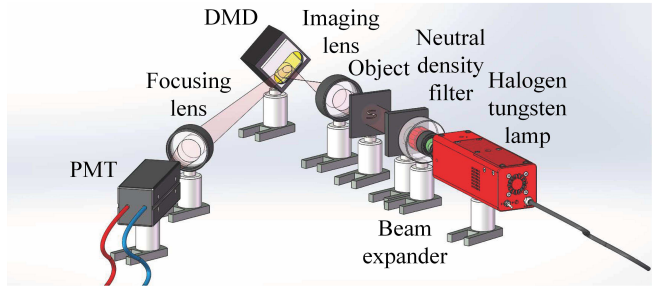}
\caption{Schematic of the experimental setup. After being collimated and broadened, the light illuminates the object and the DMD, and the total photon counts are recorded by a PMT.}
\label{fig:exp}
\end{figure}

The carrier generation process of our proposal is based on a photon-counting single-pixel camera, as presented in figure~\ref{fig:exp}. The binary modulation patterns are provided by a DMD. In the experiment, the thermal light from a stabilized halogen tungsten light source with a wavelength range from 360~nm to 2600~nm passes through a beam expander and a neutral density filter (NDF) onto a black-and-white film (the object) with a printed transparent letter ``S''. Here the NDF is used to attenuate the light to the ultra-weak light level. The object is then imaged directly onto the DMD which consists of $1024\times768$ micromirrors. Each micromirror is switched between two positions oriented at $\pm12^\circ$, which correspond to a bright pixel 1 and a dark pixel 0, determined by the pattern. The reflected light in one of the orientations is converged onto one spot by a focusing lens and is collected by a counter-type Hamamatsu H10682-210 PMT. Since the PMT counts the number of the photons that is proportional to the total light intensity for each pattern, it can be treated as a single-photon detector. Actually, the recorded photon sequence $y$ is a measured signal of the object. Following the procedures given in figure~\ref{fig:watermark}, it is simple for the receiver to obtain the signal $y'''$ from the public network, to extract the watermark, and to recover the signal $b$ according to the random allocation keys 1 and 2. It is worth noting that all the values of $y'$ should be converted into a binary sequence for XOR operation, and $b''$ should be turned back to a decimal sequence at last. Using the measurement matrix, the image is recovered by total variation minimization for identity authentication. Here the experimental setup is simple and the whole techniques reduce the complexity.
\begin{figure}[htbp]
\centering
\includegraphics[width=0.86\linewidth]{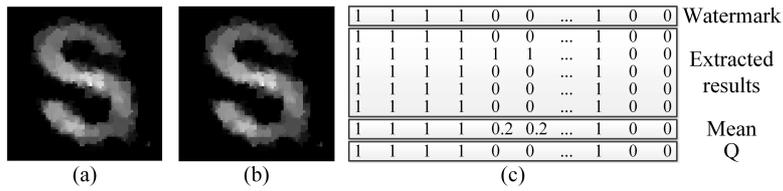}
\caption{Experimental results. (a) and (b) are obtained from the measured signal $y$ and the calculated signal $b$, respectively, but with the same matrix $A$. The number of measurements $M$ (i.e., the length of the signal) for (a) and (b) is both 2560. Here the watermark is also reused five times, with each watermark's size of 16~bits (2~bytes), thus the signal can be distributed into 80~groups (i.e., $r=32$~bits). The BER of the watermark is 0.00\%, and the PSNR of (b) refers to (a) is 44.09~dB. (c) is the intermediate process for the watermark extraction.}
\label{fig:expresults}
\end{figure}

Each pattern corresponds to one row of the measurement matrix $A$, and has $64\times64$ pixels, of the same size as the CS image. There are $M$ binary patterns, thus the size of $A$ is $M\times N$ with $N=64\times64$. The experimental results [figure~\ref{fig:expresults}] are in accord with figure~\ref{fig:simulation}. In the case of five-repetition usage of the same watermark, we get five sets of extracted results, compute their average, and then set the value to 0 if the mean is less than 0.5, and 1 otherwise. From figure~\ref{fig:expresults}(c), it can be seen that the calculated watermark is accurate with large probability. Here $A$ is generated by a random physical process. Since the photon fluctuations contain the stochastic intensity fluctuations of the light source, scattering noise along the light path, and shot noise, the signal $y$ has true randomness as well as good information imperceptibility. Here the matrix $A$, the random allocation keys 1 and 2, and some other essential parameters (like the imaging area, the pixel size, the number of measurements, the allocation selection strategy) are all used as the initial keys, which are distributed to the legitimate parties through absolutely secure approaches (e.g., secure hash algorithm 3, quantum key distribution protocol) in advance. In the future, we can also make a difference between each adjacent frames to perform the ``positive-negative'' modulation and acquire a very satisfactory image quality \cite{YuSR2014}. For full-colour imaging, one can use three spectral filters to restore the red, green, and blue sub-images, extending the dimension of the signal carrier.

\section{Security analysis and discussion}
\begin{figure}[htbp]
\centering
\includegraphics[width=0.86\linewidth]{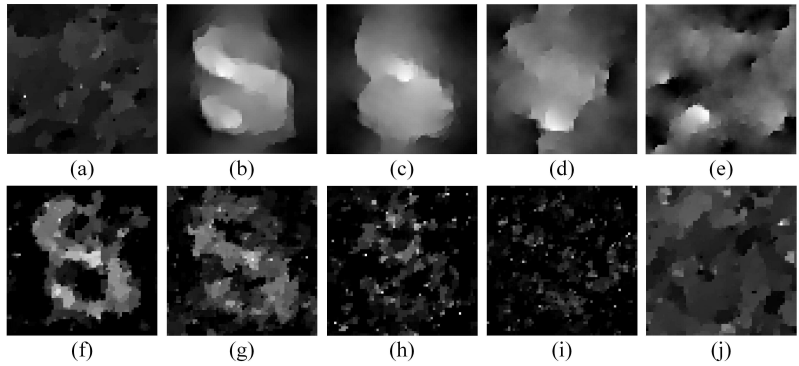}
\caption{Reconstructions after some attacks. The original recovered image that corresponds to the unassailed signal $y'''$ can be found in figure~\ref{fig:expresults}(b). Since $y'''$ is transmitted through the public network, the attacker/eavesdropper Eve can arbitrarily tamper with the data. (a) is a CS image that is recovered with $y'''$ completely disordered by Eve, and with a PSNR of 13.68~dB and a BER of 56.25\%. Then, take any value of the signal $y'''$ for example, make the $13^{\textrm{th}}$, $14^{\textrm{th}}$, $15^{\textrm{th}}$, $16^{\textrm{th}}$ bit (counting from low to high) incorrect, we will obtain the results (b)--(e), with a PSNR of 12.81~dB, 11.70~dB, 11.45~dB, 10.89~dB, respectively. Since the value is stored in doubles type, i.e., each element takes up 64~bits (8~bytes), our method can detect this kind of tampering with a probability of $52/64=81.25\%$, for the case that only one bit has been falsified. (f)--(i) are the retrieved images with 8, 16, 32, 64 values of $y'''$ substituted with 0, and with a PSNR of 18.08~dB, 16.88~dB, 14.27~dB, 13.32~dB, respectively. (j) is the reconstructed image with one value missing and all the subsequent data is shifted. Its PSNR is 12.82~dB and BER is 37.50\%. Here the length of $y'''$ is all 2560, and the watermark is also reused five times.}
\label{fig:attack}
\end{figure}
Now let us discuss the security of our one-time-pad protocol against different attacks. Here, for simplicity and without loss of generality, we directly perform our security analysis on the experimentally recorded signal. Assume that the eavesdropper Eve possesses any advanced technique. In order to ensure information transmission security, further channel coding and decoding can be used. Even if Eve knows complete $y'''$ from the public network, to recover the watermark she must guess the random allocation key 2 exactly, which costs a full permutation of $M$. In the case of figure~\ref{fig:attack}, we have $M=2560$, thus Eve must sweep through $M!\gg2^M\gg M^2$ (when $M>=5$) possible permutations. For example, from $170!=7.2574\times10^{306}$, we see that $2560!$ is thought to be extremely large. Assuming that each run costs a few milliseconds for a supercomputer, it will take many years to finish the exhaustion. Furthermore, if Eve wants to get the CS image, she must firstly spend another $M!$ possible permutations to guess the random allocation key 1 and recover the signal $b$, and secondly guess all the frames and secure parameters. For a $2560\times4096$ binary measurement matrix, where $4096=64\times64$, the probability of it being deciphered through exhaustive attack is $2^{10485760}$. Actually, the size of the image and the measurement matrix can be much larger, and the measurement matrix owned by each user can also be different. So it is impossible for her to decipher the communication.

Additionally, Eve can also intercept the signal $y'''$, falsify some bits in it and then send it to the legitimate users, as illustrated in figure~\ref{fig:attack}. However, the legitimate party can easily discover Eve's presence by checking the quality of recovered images, whether she completely disorders the signal $y'''$, or only changes one fixed bit in each value, or takes some values away. If one user finds that the quality of CS image is poor, then he/she can conclude that there must exist an attack. Therefore, it is demonstrated that our method has a highly sensitivity to the intentional tampering and an outstanding performance for data tamper detection. It is worth mentioning that our protocol is robust against the attack because of its particular scheme, rather than the experiment. Thus we can regard this protocol as being greatly safe for practical applications in defense, communication, and financial market.

\section{Conclusions}
In conclusion, we propose a secure protocol for cryptographic key distribution over a public network, where a watermark is treated as the cryptographic key to be distributed from the server to the valid parties. The compressive sampling is experimentally performed via a single-pixel photon-counting imaging system, in which the photon-limited measurements contribute additional security of the protocol. The watermark is embedded directly in the randomly-rearranged compressive measurements, rather than in the image or its non-overlapping divided blocks, ensuring the confidentiality and imperceptibility of the cryptographic key and avoiding the content leakage. Thus the watermark embedding is actually performed between compressive sampling and authentication procedures. At the receiving terminal, based on the initial keys and the variance of disordered measurement sequence that is send through the public network, the legitimate users have full access to both the accurate watermark and the encrypted image, the latter can be used for efficient authentication and tampering identification. Both numerical simulation and experimental results show that our approach owns a good robustness against different eavesdropping attacks. This protocol may pave the way for applying the variance characteristic of the measurements in optical secure communication, and will have a great impact on many practical applications of optical security.

\section*{Acknowledgements}
The author greatly appreciates Ning Wu for offering help in improving the manuscript. This work is supported by the National Natural Science Foundation of China [grant number 61801022], the Beijing Natural Science Foundation [grant number 4184098], the National Key Research and Development Program of China [grant number 2016YFE0131500], the International Science and Technology Cooperation Special Project of Beijing Institute of Technology [grant number GZ2018185101], and the Beijing Excellent Talents Cultivation Project - Youth Backbone Individual Project.

\end{document}